\documentclass[manuscript,superscriptaddress]{emulateapj}
\usepackage{graphicx}
\usepackage{amssymb}
\usepackage{amsmath}
\usepackage{textcomp}

\newcommand{\Msun}{M\ensuremath{_{\odot}}}
\newcommand{\Zsun}{Z\ensuremath{_{\odot}}}
\newcommand{\Hcc}{cm\ensuremath{^{-3}}}
\newcommand{\eos}{EoS}
\newcommand{\eospc}{PC}
\newcommand{\eosss}{SS}
\newcommand{\pdf}{PDF}
\newcommand{\mw}{MW}
\newcommand{\mwpc}{MW\ensuremath{_{\mathrm{PC}}}}
\newcommand{\smc}{SMC}
\newcommand{\smcZth}{SMC\ensuremath{_{0.1{\mathrm{Z}}_{\odot}}}} 
\newcommand{\smcZrd}{SMC\ensuremath{_{0.3{\mathrm{Z}}_{\odot}}}}
\newcommand{\smcZsun}{SMC\ensuremath{_{1.0{\mathrm{Z}}_{\odot}}}}
\newcommand{\smcpc}{SMC\ensuremath{_{\mathrm{PC}}}}
\newcommand{\smcss}{SMC\ensuremath{_{\mathrm{SS}}}}
\newcommand{\lmc}{LMC}
\newcommand{\lmcZsun}{LMC\ensuremath{_{1.0{\mathrm{Z}}_{\odot}}}}
\newcommand{\lmcpc}{LMC\ensuremath{_{\mathrm{PC}}}}
\newcommand{\lmcss}{LMC\ensuremath{_{\mathrm{SS}}}}
\newcommand{\SigmaSFR}{\ensuremath{\Sigma_{\mathrm{SFR}}}}
\newcommand{\Sigmagas}{\ensuremath{\Sigma_{\mathrm{gas}}}}

\begin{document}
\title{The role of turbulence in star formation laws and thresholds}
\author{Katarina Kraljic}
\author{Florent Renaud}
\author{Fr\'ed\'eric Bournaud}
\affiliation{CEA, IRFU, SAp, F-91191 Gif-sur-Yvette Cedex, France}
\author{Fran\c coise Combes}
\affiliation{Observatoire de Paris, LERMA et CNRS, 61 Av de l'Observatoire, F-75014 Paris, France}
\author{Bruce Elmegreen}
\affiliation{IBM T. J. Watson Research Center, 1101 Kitchawan Road, Yorktown Heights, New York 10598 USA}
\author{Eric Emsellem}
\affiliation{European Southern Observatory, 85748 Garching bei Muenchen, Germany}
\affiliation{Universit\'e Lyon 1, Observatoire de Lyon, CRAL et ENS, 9 Av Charles Andr\'e, F-69230 Saint-Genis Laval, France}
\author{Romain Teyssier}
\affiliation{Institute for Theoretical Physics, University of Z\"urich, CH-8057 Z\"urich, Switzerland}

\begin{abstract}
The Schmidt-Kennicutt relation links the surface densities of gas to the star formation rate in galaxies. The physical origin of this relation, and in particular 
its break, i.e. the transition between an inefficient regime at low gas surface densities and a main regime at higher densities, remains debated. Here, we study 
the physical origin of the star formation relations and breaks in several low-redshift galaxies, from dwarf irregulars to massive spirals. We use numerical simulations 
representative of the Milky Way, the Large and the Small Magellanic Clouds with parsec up to subparsec resolution, and which reproduce the observed star formation 
relations and the relative variations of the star formation thresholds. We analyze the role of interstellar turbulence, gas cooling, and geometry in drawing 
these relations, at 100 pc scale. We suggest in 
particular that the existence of a break in the Schmidt-Kennicutt relation could be linked to the transition from subsonic to supersonic turbulence
and is independent of self-shielding effects. This transition being connected to the gas thermal properties and thus to the metallicity, the break is shifted 
toward high surface densities in metal-poor galaxies, as observed in dwarf galaxies. 
Our results suggest that together with the collapse of clouds under self-gravity, turbulence (injected at galactic scale) can induce the compression of 
gas and regulate star formation.
\end{abstract}


\section{Introduction}

Star formation is among the most important physical processes affecting the formation and evolution of galaxies. 
Nevertheless, fundamental questions about the efficiency of the conversion of gas into stars and what triggers the process of star formation remain open.

Observations of galaxies have shown a close correlation, known as the Schmidt-Kennicutt relation, between the surface density of star formation rate (\SigmaSFR) and 
the surface density of gas (\Sigmagas) \citep[e.g.][]{Kennicutt1989, WongBlitz2002}. 
The details of this scaling relation are found to vary with the environment. 
Spiral galaxies convert their gas into stars with longer depletion times than galaxies in a merger phase \citep{Daddi2010, Genzel2010, Saintonge2012}, 
but more rapidly than dwarf galaxies \citep{Leroy2008,Bolatto2011}. 
In addition, the observed relation varies for different tracers. It is shallower for molecular gas than for total (molecular and atomic) 
gas \citep{GaoSolomon2004, Bigiel2008, Heyer2004}, but steeper when the atomic gas is considered solely
\citep{Kennicutt1998, Kennicutt2007, Schuster2007, Bigiel2008}. Several other observations \citep[e.g.][]{Kennicutt1989, MartinKennicutt2001, Boissier2003} 
have suggested the existence of a critical surface density, the so-called break, below which \SigmaSFR{} in spiral galaxies drops: star formation 
is inefficient compared to the regime at high surface densities, well described by a power-law.  
Consequently, a composite relation of star formation seems to be a better description for the \Sigmagas--\SigmaSFR{} relation, rather than a single power-law.

However, the origin of the break, and the transition to a different regime of star formation at high surface densities, remain a matter of debate. 
Some models evoke the Toomre criterion for gravitational instability \citep[e.g.][]{Quirk1972, Kennicutt1989, MartinKennicutt2001} or 
rotational shear \citep{Hunter1998,MartinKennicutt2001} to interpret the existence of the break.
\cite{Elmegreen1994} emphasize the need for the coexistence of two thermal phases in pressure equilibrium and  
\cite{Schaye2004} argues that it is the transition from the warm to the cold gas phase, enhanced by the ability of the gas to shield itself from external
photo-dissociation, that triggers gravitational instabilities over a wide range of scales. 
Self-shielding plays an important role also in the model of \cite{Krumholz2009}, where it sets the transition from atomic to molecular phase at a 
metallicity-dependent \Sigmagas. 
\cite{DibPMB2011} shows that feedback from massive stars is an important regulator of the star formation efficiency in protocluster forming clouds. 
Based on this, \cite{Dib2011} proposes that the break in the \Sigmagas--\SigmaSFR{} relation can be related to a feedback-dependent transition of the star
formation efficiency per unit time, as a function of the gas surface density (but see \citealt{Dale2013} reporting a possibly low impact of the stellar feedback
on the star formation rate and efficiency).
\cite{Renaud2012} have recently proposed an analytic model in which the origin of the star formation break
is related to the turbulent structure of the interstellar medium (ISM). In this model, a threshold in the local volume density, resulting in the observed
surface density break corresponds 
to the onset of supersonic turbulence that generates shocks which in turn trigger the gravitational instabilities and thus star formation.

Different mechanisms are invoked in theoretical works to explain the scaling relations, such
as gravity \citep{Tan2000, SilkNorman2009}, turbulence of the interstellar medium \citep[e.g.][]{Elmegreen2002, MacLowKlessen2004, Krumholz2005, Hennebelle2011,
Padoan2011, Federrath2012, Renaud2012, Federrath2013}, feedback from massive stars \citep{Dib2011} and the interplay between the dynamical and thermal state 
of the gas \citep{Struck1999}.

In addition to these theoretical studies, several galaxy simulations modeling the ISM found a reasonable agreement with observations, using various recipes for star formation and stellar feedback 
\citep{Li2006, WadaNorman2007, RobertsonKravtsov2008, TaskerBryan2008, DobbsPringle2009, KoyamaOstriker2009, Agertz2011, Dobbs2011, Kim2011, Monaco2012,
Rahimi2012, ShettyOstriker2012,Halle2013}. Among them \cite{Bonnell2013} resolved the small scale physics of star formation in the context of galactic scale 
dynamics. 
The observed correlation between \Sigmagas{} and \SigmaSFR, together with the break of \SigmaSFR{} are often reproduced in simulations, 
but it remains unclear to what extent the star formation rate estimates depend on the 
parameters of individual models and underlying assumptions, and what are the fundamental drivers for the observed relations.

In this paper we aim at providing a better understanding of the star formation relations and threshold by studying the local properties of simulated galaxies.
Our work is in great part motivated by the analytic model of \cite{Renaud2012}, based on the supersonic nature of the turbulence in the ISM. 
Their formalism leads to an analytic expression relating \Sigmagas{} and \SigmaSFR{} that depends on three parameters: 
the Mach number, the star formation density threshold and the thickness of the star-forming regions.  
One assumption of this model is the characterization of an entire star-forming region by a single set of these three parameters, while wide ranges of them 
are more appropriate for describing the real ISM. Another assumption is the description of the gas volume density by a log-normal distribution, which was 
primarily found for isothermal supersonic turbulence \citep[e.g.][]{Vazquez1994,NordlundPadoan1999}.
By performing galaxy simulations, we achieve a wide diversity of parameters and density distributions consistent with the multiphase ISM. 
Simulations allow us to study local properties of individual regions of galaxies, such as velocity dispersion, temperature or geometry and to infer their 
impact on the \Sigmagas{} -- \SigmaSFR{} relation.

We start with the presentation of the simulation technique and details of galaxy models in Section \ref{Sec:Simulations}. Section \ref{Sec:3galaxies} presents
the analyzed sample of galaxies. The method used in deriving parameters needed for the analysis is described in Section \ref{Sec:Analysis}.
The dependence of star formation on different parameters, plotted in the \Sigmagas{} -- \SigmaSFR{} parameter space, is shown in Section \ref{Sec:Results}.
In Section \ref{Sec:Discussion}, we discuss the results and compare with the model of \cite{Renaud2012}. Finally, we conclude with the summary in 
Section \ref{Sec:Summary}.


\section{Simulation technique}
\label{Sec:Simulations}

We use the Adaptive Mesh Refinement (AMR) code RAMSES \citep{Teyssier2002} to model a set of isolated galaxies, as in \cite{Renaud2013}.
Physical parameters used here are described in Section~\ref{Sec:3galaxies}.

The dark matter and stellar components are evolved using a particle-mesh solver.
The dynamics of the gaseous component is computed by solving hydrodynamics equations on the adaptive grid using a second-order Godunov scheme.

The refinement strategy for all our simulations is based on the density criterion of stars and gas.  
In order to account for the unresolved physics due to finite resolution, the so-called Jeans polytrope (T $\propto \rho$) is added at 
high densities, corresponding to the scales smaller than the maximal resolution. This additional thermal support ensures that the thermal Jeans length is 
always resolved by at least four cells and thus avoids numerical instabilities and artificial fragmentation \citep{Truelove1997}.

\subsection{Star formation}
\label{Subsec:SF}
During the simulations, stellar particles are formed by conversion of gas. These particles are used for the injection of stellar feedback, but are not
used in the post processing analysis of the SFR, which is recalculated from the density of gas (see Section \ref{Sec:Analysis}).

Details of star formation and the associated stellar feedback are given in \cite{Renaud2013}.
The values of the star formation efficiency $\epsilon$ and the star formation threshold density $\rho_0$ that we have adopted here 
(see Table \ref{Tab:simulations}) are adjusted to match the observed global SFR
for local galaxies: $\approx 1-5$ \Msun yr$^{-1}$ for the Milky Way \citep{Robitaille2010} and $\approx 0.4$ \Msun yr$^{-1}$ for the Large Magellanic 
Cloud \citep[e.g.][]{Skibba2012}, on average. SFR of the simulated Small Magellanic Cloud is $\approx 0.5$ \Msun yr$^{-1}$, on average, which is higher
than the observed value (e.g., 0.05 \Msun yr$^{-1}$ obtained by \citealt{Wilke2004}), perhaps because of different structures, but a one-to-one match is not 
seeked.

The stellar feedback is modeled by photo-ionization together with radiative pressure in the case of the Milky Way simulation \citep{Renaud2013} and supernova (SN) 
feedback, implemented as a Sedov blast \citep{DuboisTeyssier} in all simulations, either in a kinetic or thermal scheme.
The choice of the SN feedback scheme is determined by treatment of the heating and cooling processes (see Section \ref{Sec:EOS}): every time the gas follows
an equation of state (\eos), the total energy of SN ($10^{51}$ erg) is injected in the kinetic form, since thermal feedback would have no effect.

\subsection{Metallicity and Equation of state}
\label{Sec:EOS}

The cooling and heating processes occurring in the ISM depend on the metallicity.
In our models, the gas cooling due to atomic and fine-structure lines, and radiation heating from an uniform ultraviolet  
background are modeled following \cite{Courty} and \cite{HaardtMadau1996}, respectively.
Metallicity is a parameter fixed for each simulation and represents the average metal mass fraction in the galaxy.

\medskip

\begin{figure}[h!]
\begin{center}
\includegraphics[width=\columnwidth]{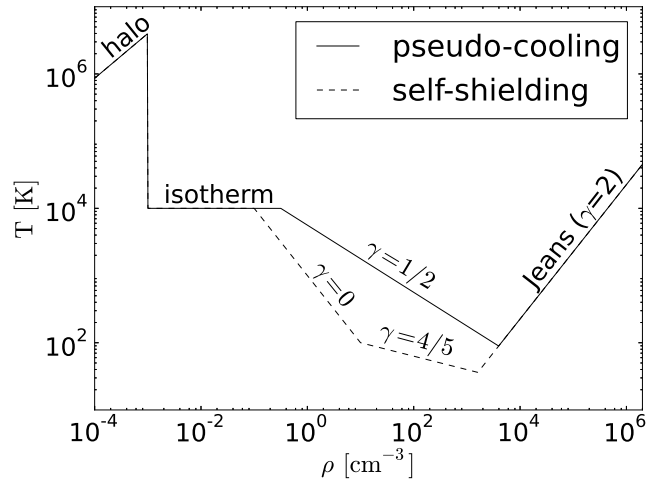}
\caption{\label{Fig:eos_pc_ss} Effective \eos{} for pseudo-cooling (solid line) and self-shielding (dashed line). 
Pseudo-cooling \eos{} mimics detailed balance between cooling and heating processes at 1/3 solar metallicity.
Self-shielding \eos{} models the ability of the gas clouds to reach even lower temperatures by absorbing the interstellar radiation in their outer layers. 
The low density regime of index $\gamma=$ 5/3 corresponds to the hot virialized gas in the stellar halo. 
The Jeans polytrope of index $\gamma=$ 2 dominates high density regions, avoiding artificial fragmentation.
For a given spatial resolution $d_{x,min}$ the Jeans polytrope becomes 
active at densities above $\approx$ 2761 \Hcc{} $\times$ ($d_{x,min}$/1 pc)$^{-4/3}$ in the case of \eospc{} 
and $\approx$ 1023 \Hcc{} $\times$ ($d_{x,min}$/1 pc)$^{-5/3}$ in the case of \eosss{}. The corresponding temperatures are 
$\approx$ 107 K $\times$ ($d_{x,min}$/1 pc)$^{2/3}$ and $\approx$ 40 K $\times$ ($d_{x,min}$/1 pc)$^{1/3}$ for \eospc{} and \eosss, respectively.
}
\end{center}
\end{figure}

Heating and cooling processes can often substantially slow down the simulation. 
A piecewise polytropic \eos{} of form $T \propto \rho^{\gamma - 1}$ with the polytropic index $\gamma$ can be applied instead.
We use a pseudo-cooling (\eospc{}) \eos{} (Figure \ref{Fig:eos_pc_ss}), fitting the heating and cooling equilibrium of gas at 1/3 solar metallicity 
\citep{Bournaud2010,TeyssierChapon2010}. In the above definition of the \eos, we neglect the capacity of the gas to shield itself from the surrounding 
radiation. At densities around 0.1 -- 1 \Hcc{} and temperatures of several hundreds K, self-shielding (\eosss) becomes important: the molecular fraction of the 
gas increases, enabling it to cool down to even lower temperatures \citep{Dobbs2008}. This can be modeled by the alternative \eos{} shown in 
Figure \ref{Fig:eos_pc_ss}.

\section{Galaxy sample}
\label{Sec:3galaxies}

\subsection{Initial conditions}
We study models of a spiral galaxy resembling the Milky Way (hereafter \mw{}), a disc galaxy similar to the Large Magellanic Cloud (\lmc{}) and an
irregular dwarf galaxy comparable to the Small Magellanic Cloud (\smc{}). We do not try to reproduce fine details for these galaxies, but propose
models representing systems with different morphological and physical properties. Each simulation is performed in isolation and without cosmological evolution.

The details of the \mw{} simulation can be found in \cite{Renaud2013}. Here, this simulation is analyzed at resolution comparable to the resolution of
other galaxy simulations in our sample, which is 1.5 pc, i.e. not at its maximal resolution.
The parameters of all simulations are summarized in Table~\ref{Tab:simulations}. 

Simulations are labeled in a way to stress their principal difference which is related to \eos{} or metallicity parameter. Simulations in which the 
heating and cooling processes are evaluated have the value of metallicity in subscript. 
If the \eos{} is used instead, the subscript indicates the name of the equation of state. The most realistic cases are 
the \lmcZsun{} and the \smcZth{} simulations for \lmc{} and \smc{}, respectively. The solar metallicity we have adopted in the \lmcZsun{} simulation is higher 
than in the real \lmc{} \citep[1/2 \Zsun{};][]{Russell1992,Rolleston1996}, but fairly representative of low-redshift and low mass disc galaxy that we intend 
to model. The metallicity of 1/10 \Zsun{} that we used in the \smcZth{} simulation falls in the range of estimated values for the real \smc{} 
\citep[1/5--1/20 \Zsun{};][]{Russell1992,Rolleston1999}.

\begin{table*}
\begin{center}
\caption{\label{Tab:simulations} Summary of model parameters.}
\begin{ruledtabular}
\renewcommand{\arraystretch}{1.5}
\begin{tabular}{c c  c c c c c c c c c}
  \multicolumn{1}{c|}{ }& \multicolumn{1}{c|}{\mwpc\footnote{simulations are labeled mnemonically, with their name having the value of the metallicity 
   or \eos{} parameter in subscript: \mwpc, \lmcZsun, \lmcpc, \lmcss, \smcZth, \smcZrd, \smcZsun, \smcss, \smcpc}} 
   & \multicolumn{1}{c}{\lmcZsun} & \multicolumn{1}{c}{\lmcpc} & \multicolumn{1}{c|}{\lmcss} &  \multicolumn{1}{c}{\smcZth}
   & \multicolumn{1}{c}{\smcZrd} & \multicolumn{1}{c}{\smcZsun} & \multicolumn{1}{c}{\smcpc} & \multicolumn{1}{c}{\smcss}\\
  \hline 
  \multicolumn{1}{c|}{\eos{} or metallicity\footnote{metallicity is a meaningful parameter only when the heating and cooling processes are evaluated, the name 
  of the \eos{} is given otherwise} [\Zsun]} & 
  \multicolumn{1}{c|}{\eospc{}} & \multicolumn{1}{c}{1.0} & \multicolumn{1}{c}{\eospc{}} &\multicolumn{1}{c|}{\eosss{}} & \multicolumn{1}{c}{0.1} &
  \multicolumn{1}{c}{0.3} & \multicolumn{1}{c}{1.0} & \multicolumn{1}{c}{\eospc{}} &  \multicolumn{1}{c}{\eosss{}}\\ 
  \hline
  \multicolumn{1}{c|}{box length [kpc]} & \multicolumn{1}{c|}{100} & \multicolumn{3}{c|}{25} & \multicolumn{5}{c}{30}\\
  \multicolumn{1}{c|}{AMR coarse level} & \multicolumn{1}{c|}{9} & \multicolumn{3}{c|}{8} & \multicolumn{5}{c}{8}\\
  \multicolumn{1}{c|}{AMR fine level} & \multicolumn{1}{c|}{21} & \multicolumn{3}{c|}{14} & \multicolumn{5}{c}{15}\\
  \multicolumn{1}{c|}{maximal resolution [pc]} & \multicolumn{1}{c|}{0.05\footnote{the analysis is performed by extracting the simulation data at the effective resolution of 1.5 pc (see text for details)}} & \multicolumn{3}{c|}{1.5} & \multicolumn{5}{c}{1.0}\\
  \hline
  \multicolumn{1}{c|}{DM halo} & \multicolumn{1}{c|}{} & \multicolumn{3}{c|}{} & \multicolumn{5}{c}{}\\
  \multicolumn{1}{c|}{mass [$\times$ $10^9$ \Msun]} & 
  \multicolumn{1}{c|}{453.0} & \multicolumn{3}{c|}{8.0} & \multicolumn{5}{c}{1.2}  \\ 
  \multicolumn{1}{c|}{number of particles [$\times$ $10^5$]} &
  \multicolumn{1}{c|}{300.0} & \multicolumn{3}{c|}{3.49} & \multicolumn{5}{c}{5.0}  \\ 
  \hline
  \multicolumn{1}{c|}{primordial stars\footnote{stars initially present in simulation}} & \multicolumn{1}{c|}{} & \multicolumn{3}{c|}{} & \multicolumn{5}{c}{}\\
  \multicolumn{1}{c|}{mass [$\times$ $10^9$ \Msun]} & 
  \multicolumn{1}{c|}{46.0} & \multicolumn{3}{c|}{3.1} & \multicolumn{5}{c}{0.35}  \\ 
  \multicolumn{1}{c|}{number of particles [$\times$ $10^5$]} &
  \multicolumn{1}{c|}{300.0} & \multicolumn{3}{c|}{5.75} & \multicolumn{5}{c}{2.15}  \\ 
  \hline
  \multicolumn{1}{c|}{gas} & \multicolumn{1}{c|}{} & \multicolumn{3}{c|}{} & \multicolumn{5}{c}{}\\ 
  \multicolumn{1}{c|}{mass [$\times$ $10^9$ \Msun]} & 
  \multicolumn{1}{c|}{5.94} & \multicolumn{3}{c|}{0.54} & \multicolumn{5}{c}{0.715}  \\ 
  \multicolumn{1}{c|}{$\sim$ AMR cell number [$\times$ 10$^6$]} & \multicolumn{1}{c|}{240} & \multicolumn{1}{c}{385} &
  \multicolumn{1}{c}{440} & \multicolumn{1}{c|}{450} & \multicolumn{1}{c}{43} & \multicolumn{1}{c}{43} & \multicolumn{1}{c}{43} & \multicolumn{1}{c}{45} 
  & \multicolumn{1}{c}{50} \\
  \multicolumn{1}{c|}{radial profile} & \multicolumn{1}{c|}{exponential} & \multicolumn{3}{c|}{exponential} & 
  \multicolumn{5}{c}{exponential}\\ 
  \multicolumn{1}{c|}{scale radius [kpc]} & 
  \multicolumn{1}{c|}{6} & \multicolumn{3}{c|}{3} & \multicolumn{5}{c}{1.3}  \\ 
  \multicolumn{1}{c|}{radial truncation [kpc]} & 
  \multicolumn{1}{c|}{28} & \multicolumn{3}{c|}{6} & \multicolumn{5}{c}{2.3}  \\ 
  \multicolumn{1}{c|}{vertical profile} & \multicolumn{1}{c|}{exponential} & \multicolumn{3}{c|}{exponential} & 
  \multicolumn{5}{c}{exponential}\\ 
  \multicolumn{1}{c|}{scale-height [kpc]} & 
  \multicolumn{1}{c|}{0.15} & \multicolumn{3}{c|}{0.15} & \multicolumn{5}{c}{0.6}  \\ 
  \multicolumn{1}{c|}{vertical truncation [kpc]} & 
  \multicolumn{1}{c|}{1.5} & \multicolumn{3}{c|}{0.45} & \multicolumn{5}{c}{1.3}  \\ 
  \hline
  \multicolumn{1}{c|}{intergalactic density\footnote{fraction of density of gas at the edge of galaxy that is set beyond the truncation of the gas disc}} & 
  \multicolumn{1}{c|}{$10^{-7}$} & \multicolumn{3}{c|}{$10^{-7}$} & \multicolumn{5}{c}{$10^{-3}$}  \\ 
  \hline
  \multicolumn{1}{c|}{star formation} & \multicolumn{1}{c|}{} & \multicolumn{3}{c|}{} & \multicolumn{5}{c}{}\\ 
  \multicolumn{1}{c|}{$\epsilon$} & \multicolumn{1}{c|}{$3\%$} &
  \multicolumn{3}{c|}{$3\%$} & \multicolumn{5}{c}{$3\%$}\\ 
  \multicolumn{1}{c|}{$\rho_0$ [\Hcc]} & \multicolumn{1}{c|}{$2\times 10^3$} & \multicolumn{3}{c|}{$10^2$} &
  \multicolumn{5}{c}{$10^2$}\\ 
  \hline
  \multicolumn{1}{c|}{stellar feedback} & \multicolumn{1}{c|}{} & \multicolumn{3}{c|}{} & \multicolumn{5}{c}{}\\ 
  \multicolumn{1}{c|}{photo-ionization} & \multicolumn{1}{c|}{\checkmark} & 
  \multicolumn{3}{c|}{\textminus} & \multicolumn{5}{c}{\textminus} \\ 
  \multicolumn{1}{c|}{radiative pressure} & \multicolumn{1}{c|}{\checkmark} & \multicolumn{3}{c|}{\textminus} &
  \multicolumn{5}{c}{\textminus} \\ 
  \multicolumn{1}{c|}{SNe} & \multicolumn{1}{c|}{kinetic} & \multicolumn{1}{c}{thermic} & \multicolumn{1}{c}{kinetic} & 
  \multicolumn{1}{c|}{kinetic} &  \multicolumn{1}{c}{thermic} & \multicolumn{1}{c}{thermic} & \multicolumn{1}{c}{thermic} & \multicolumn{1}{c}{kinetic} & \multicolumn{1}{c}{kinetic}\\
\end{tabular}
\end{ruledtabular}
\end{center}
\end{table*}

\subsection{Morphology}
\begin{figure*}[t!]
\begin{center}
\includegraphics[width=\textwidth]{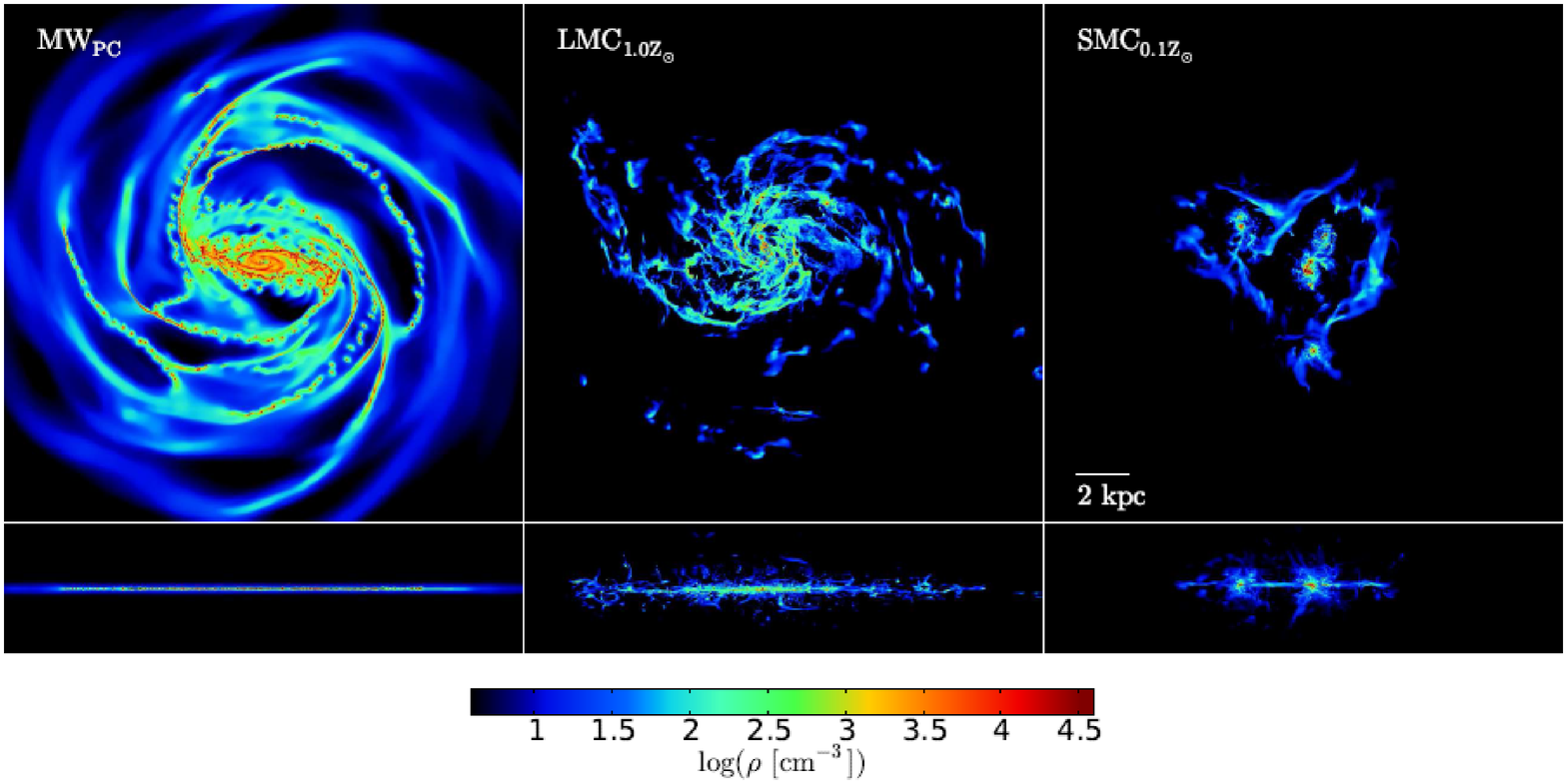}
\caption{\label{Fig:3maps} Surface density of gas of the galaxies for our three simulations: \mwpc{} (left), \lmcZsun{} (middle) and \smcZth{} 
(right panel) seen face-on on the top and edge-on on the bottom panels. The box size of the face-on projection is 20$\times$20 kpc$^2$ and that of the 
edge-on projection is 20$\times$5 kpc$^2$.}
\end{center}
\end{figure*}

Figure \ref{Fig:3maps} displays the surface gas density map of the three galaxies. 
\mwpc{}, a spiral galaxy, shows large variety of substructures: bar and spiral arms on the kpc-scale as well as dense clumps on the 
parsec scale \citep[see][for details]{Renaud2013}. 
\lmcZsun{} is also a disc galaxy, but with a much less pronounced structure of spiral arms and more diffuse gas present in the inter-arm regions compared to
\mwpc. \smcZth{} is an irregular dwarf galaxy. Three major dense clumps can be seen within the irregular structure of the diffuse gas.

\subsection{Gas density \pdf}
\label{Subsec:pdf}
\begin{figure}[h!]
\begin{center}
\includegraphics[width=\columnwidth]{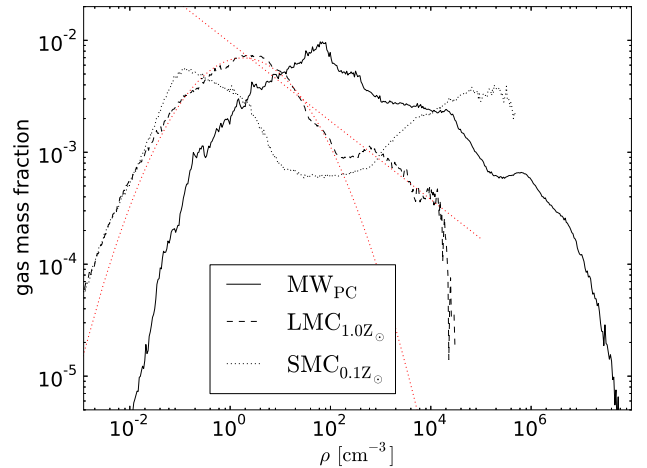}
\caption{\label{Fig:3pdf} Comparison of the mass-weighted density \pdf{} in \mwpc{} at full resolution of 0.05 pc (solid line), \lmcZsun{} (dashed line) 
and \smcZth{} (dotted line). Approximated log-normal functional form and a power-law are shown for \lmcZsun{} in red \citep[see][Figure 10, for 
the best fit for \mwpc{}]{Renaud2013}.}
\end{center}
\end{figure}

The mass-weighted density probability distribution function (\pdf) of the gas for \mwpc, \lmcZsun{} and \smcZth{} is shown in Figure \ref{Fig:3pdf}. 
The \mwpc's \pdf{} has a log-normal shape, followed by a power-law tail at high densities ($\rho\gtrsim$ 1000 \Hcc) probed thanks to the
high resolution reached in this simulation.
Similarly, the \lmcZsun's \pdf{} can be approximated by a log-normal functional form in the density range from 10$^{-2}$ to 10$^2$ \Hcc{} with and excess 
of dense gas with respect to a log-normal fit above density of about 100 \Hcc.
Truncation possibly due to the resolution limit is visible at a density of $\sim$ 2$\times$10$^4$ \Hcc.
The \pdf{} of the \smcZth{} is rather irregular with two components, one at low densities ($\sim$ 10$^{-1}$ \Hcc) and the other one at high densities 
($\sim$ 2$\times$ 10$^4$ \Hcc). Such irregular \pdf{} reveals the density contrast between diffuse gas 
and several high density clumps.

The shape of the density \pdf{} is determined by global properties of galaxies and physical processes of their ISM.
As suggested by \cite{RobertsonKravtsov2008}, the density \pdf{} can vary from galaxy to galaxy and that of a multiphase ISM can be constructed by summing
several log-normal \pdf s, each representing approximately an isothermal gas phase. Similarly, \cite{Dib2005} showed that the \pdf{} of a bistable
two-phase medium evolves into a bimodal form with a power-law tail at the high density-end in the presence of self-gravity 
\citep[see also][]{Elmegreen2011,Renaud2013}.
However, in most cases, a single, wider log-normal functional form
is a reasonably good approximation of the \pdf{} of disk galaxies up to $\gtrsim 10^{4}$ \Hcc{} \citep[see e.g.][]{TaskerBryan2008,Agertz2009}.

Note that \smcZth, which has a lower metallicity than \lmcZsun, is able to reach the highest densities. 
Metallicity is important for cooling: the more metallic gas is more efficient at cooling the gas down and should allow reaching higher densities. 
However, we do not observe such trend. This could indicate that factors other than thermal may be key in setting the gas distribution.  

Another possible explanation could be a mismatch between the choices of threshold density $\rho_0$ for star formation and the metallicity in 
the \lmcZsun. If $\rho_0$ is chosen to be low, stars will form in an intermediate density medium, i.e. without the need of gravitational collapse of a cloud into
a dense region.
Furthermore, stellar feedback helps the destruction of the densest clumps which produces more intermediate-density gas and further prevents the 
gravitational contraction leading to high densities. The maximum density of the ISM is thus lower than with a high $\rho_0$ and the resulting \pdf{} does not yield the 
classical high density power-law tail. However, in the case of \lmcZsun, the transition of the gas from high ($>$10$^3$ -- 10$^4$ \Hcc) to intermediate 
densities (10 -- 10$^2$ \Hcc, just below the actual $\rho_0$) due to the feedback would 
lead to a substantial reduction in the SFR (because of the $\rho_{\mathrm{SFR}} \propto \rho^{3/2}$ used in our model, the SFR is dominated by high-density gas). 
Consequently, feedback itself would be substantially reduced. 

Another, more likely explanation is that \smcZth{} contains a much higher gas fraction compared to \lmcZsun{} (see Table \ref{Tab:simulations}) 
leading to a much lower value of Toomre parameter ($\mathrm{Q} \propto \Sigma_{\mathrm{gas}}^{-1}$) which allows \smcZth{} to reach higher densities than in 
\lmcZsun{}.

\section{Analysis}
\label{Sec:Analysis}

To study the 100 pc scale properties of individual galaxies, analyzed regions are selected by examining the face-on projections of the gas distribution. 
We then consider sub-regions (referred to as beams throughout the paper) of 100$\times$100 pc$^2$ in the galactic plane and with galaxy scale-height
along the line of sight. A study of the impact of the beam size is presented in Section \ref{Subsec:BeamSize}.

\subsection{Definitions}
\label{Sec:Definitions}

In a given beam, the effective Mach number $\mathcal{M}$ is defined as:
\begin{equation}
\label{Eq:Mach_number}
 \mathcal{M} = \frac{\sigma_{v}}{\sqrt{3}} \frac{1}{{c_s}}, 
\end{equation}
where $\sigma_{v}$ and ${c_s}$ are the mass-weighted velocity dispersion and the mass-weighted sound speed with respect to the beam, respectively, 
calculated as follows
\begin{equation}
\label{Eq:sigma_v}
 \sigma_{v} = \sqrt{\frac{\sum\limits_i m_i v_i^2}{\sum\limits_i m_i} - \left(\frac{\sum\limits_i m_i v_i}{\sum\limits_i m_i}\right)^2},
\end{equation}
and
\begin{equation}
 {c_s} = \sqrt{\frac{\sum\limits_i m_i T_i \gamma \frac{k_{\mathrm{B}}}{m_{\mathrm{H}}} }{\sum\limits_i m_i}}.
\end{equation}

Summations are done over all AMR cells in the analyzed beam and the index $i$ refers to cell related quantities: $T_i$, $m_i$ and ${v}_{i}$ are the cell temperature, 
gas mass and speed, respectively. $\gamma = 5/3$ is the adiabatic index for monoatomic gas, $m_{\mathrm{H}}$ is the mass of the hydrogen atom and $k_{\mathrm{B}}$ 
the Boltzmann constant.

An alternative to the above ``beam-based'' average could be to compute the mass-weighted $\mathcal{M}$ with a cell velocity dispersion itself calculated with respect to 
its closest cells, but we find that this does not lead to a significant difference in the results.

Temperature in the beam is computed as mass-weighted average:
\begin{equation}
\label{Eq:Temperature}
 T = \frac{\sum\limits_i m_i T_i}{\sum\limits_i m_i}.
\end{equation}

To estimate the actual thickness of the star-forming regions within each beam, we apply Gaussian fit to 1D projection of the gas density along one of the 
mid-plane axes.
The thickness is then defined as the full width at half maximum of the resulting fit. Although the estimation method of the thickness parameter is simplistic, 
the obtained values are in good agreement with visual inspection of density maps of individual star-forming regions.

Note that in our analysis we don't use the SFR computed directly in the simulation. The main reason is that the conversion of gas into stars is modeled as a
stochastic process leading to the discretization of the \SigmaSFR{} values which make the analysis difficult by 
introducing more noise.

The SFR of a beam is estimated from the gas content of each cell by
\begin{equation}
\label{Eq:SFR}
 \rho_{\star} = \epsilon \frac{\rho}{t_{\mathrm{ff}}} \propto \epsilon \rho^{3/2} \quad \text{for $\rho > \rho_{0}$},
\end{equation}
where $\rho_{\star}$ is the local star formation rate density, $\rho$ is the density of gas in the cell, $\epsilon$ is the star formation efficiency per 
free-fall time $t_{\mathrm{ff}} = \sqrt{3 \pi/(32 G \rho)}$ and $\rho_0$ is the star formation threshold.

\SigmaSFR{} is then given by
\begin{equation}
\label{Eq:sigmaSFR}
 \Sigma_{\mathrm{SFR}} = \frac{\sum\limits_i {\rho_{\star}}_i V_i}{S},
\end{equation}
where ${\rho_{\star}}_i$ and $V_i$ are the cell SFR density and volume, respectively and $S$ is the surface of the beam.
Similarly, \Sigmagas{} is 
\begin{equation}
\label{Eq:sigmagas}
 \Sigma_{gas} = \frac{\sum\limits_i \rho_i V_i}{S},
\end{equation}
with $\rho_i$ representing the cell gas density.

\subsection{Tests}
\subsubsection{Beam size effects}
\label{Subsec:BeamSize}

Our choice of the beam size is related to the adopted analytical formalism which is tightly linked to the turbulence-driven structure of the ISM.
Supersonically turbulent isothermal gas is found to be well described by a log-normal probability distribution function. However, once these hypotheses about
the state of gas are relaxed, strictly log-normal \pdf{} is not recovered. The \pdf{}s of the density field in our sample of galaxies are close to, but not exactly log-normal functional forms when all scales 
are considered (see Section \ref{Subsec:pdf}). Individual beams should be large enough to be representative samples of star-forming regions at different
evolutionary stages. 
In addition, the choice of the beam size is somehow linked to the turbulence and its cascade from large scales where the turbulence is injected, down to the small scales, 
where the energy dissipation overcomes its transfer. In order to capture the turbulent cascade, the size of the beam should not be too large compared 
to the injection scale\footnote{i.e. about the scale-height of the gas disk \citep{Bournaud2010,Renaud2013}.}, nor too small compared to the dissipation scale. In the former case, the simulation
would capture other processes than turbulence and in the latter case, turbulence would be already dissipated.

\begin{figure*}[t!]
\begin{center}
\includegraphics[width=\textwidth]{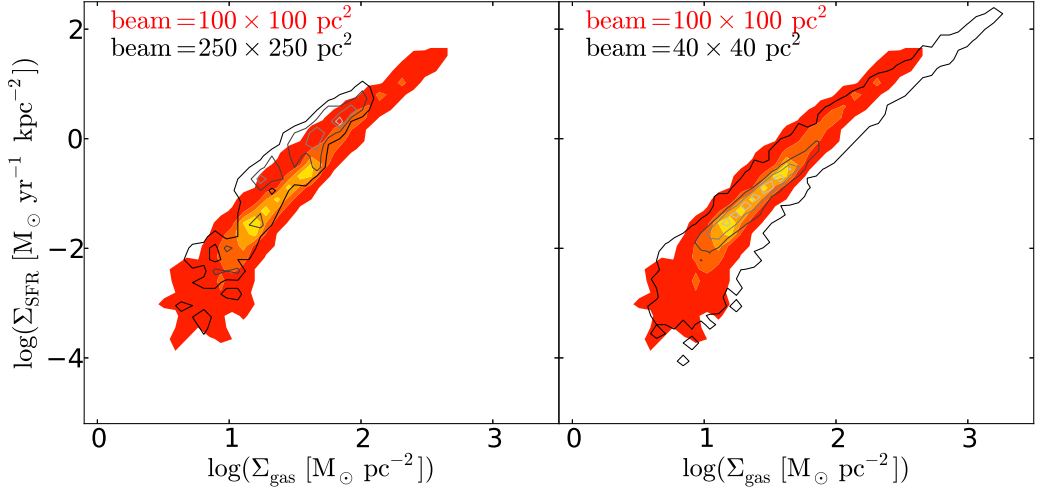}
\caption{\label{Fig:MW_beam_compare_all} The effect of beam size on the \Sigmagas--\SigmaSFR{} relation in the \mwpc{} simulation. The two panels show the 
effect of varying the beam width by a factor of 2.5 (black contours) with respect to the one used in analysis (colored filled contours). When increasing the beam size, the dense gas represents smaller and smaller volume fraction which  
leads to an overall reduction of \Sigmagas{} for the beam. The color coding of the two-dimensional normalized histogram corresponds to the 0.3, 0.5, 0.8 
and 1.0 contour levels.}
\end{center}
\end{figure*}

To estimate the impact of the size of the beam,  we compare the results in the \Sigmagas--\SigmaSFR{} plane obtained by varying the beam width by a factor of 2.5 
with respect to the one used in analysis. Figure \ref{Fig:MW_beam_compare_all} shows the comparison for the \mwpc{} simulation. Increased beam size leads to an overall
reduction of \Sigmagas{} for the beam, which can be understood as a consequence of decreased volume fraction occupied by the dense gas. On the contrary, 
smaller beam size allows reaching higher values of \SigmaSFR. 
This comparison suggests that the position of points in the \Sigmagas--\SigmaSFR{} plane depends on the considered spatial scale. \cite{Schruba2010} found such 
dependence in the study of the Local Group spiral galaxy M33. Similarly, \cite{Lada2013} found a more efficient SF at scales of molecular clouds,
indicating that caution should be used when comparing SF relations involving different spatial scales.

However, the global behavior of the \Sigmagas--\SigmaSFR{} does not seem to be strongly affected by the size of the beam, at least for the range of sizes that 
we explored.

\subsubsection{Parsec and sub-parsec physics}
\label{Sec:PcSubpcScale}

\begin{figure}[h!]
\begin{center}
\includegraphics[width=\columnwidth]{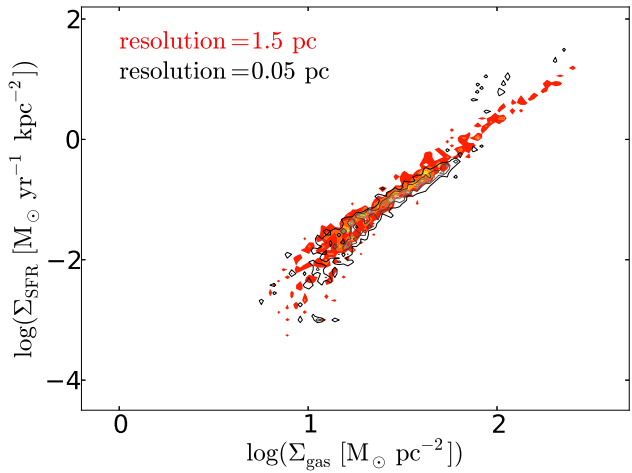}
\caption{\label{Fig:MW_res_compare} The impact of resolution on the \Sigmagas--\SigmaSFR{} relation in the \mwpc{} simulation. The maximal resolution of 0.05 pc 
(black contours) is compared to the resolution of 1.5 pc (colored filled contours), at which the entire analysis is performed. No significant difference is noticeable
at low and intermediate surface densities of gas. At high \Sigmagas, \SigmaSFR{} tends to be higher at the resolution of 0.05 pc. The color coding of
contour levels is as in Figure \ref{Fig:MW_beam_compare_all}.}
\end{center}
\end{figure}

We remind that the \mwpc{} simulation is analyzed at the resolution of 1.5 pc which is different from its maximal resolution of 0.05 pc. To study the impact of the
resolution, we compare in Figure \ref{Fig:MW_res_compare} the \Sigmagas--\SigmaSFR{} relation for these two resolutions. Sub-parsec physics
does not influence our results at low and intermediate surface gas densities, but it plays a role in densest regions, where it leads to higher values of 
\SigmaSFR. The increased resolution leads to the modification of structures mainly at high densities which translates into higher values of \SigmaSFR{} 
computed at fixed 100 pc scale.

\section{Results}
\label{Sec:Results}

In order to have a significant amount of data, we use several snapshots in the analysis of the \lmc{} and \smc{} galaxies.

\subsection{Global parameters}
\label{Sec:Global}

\begin{figure*}[t!]
\begin{center}
\includegraphics[width=\textwidth]{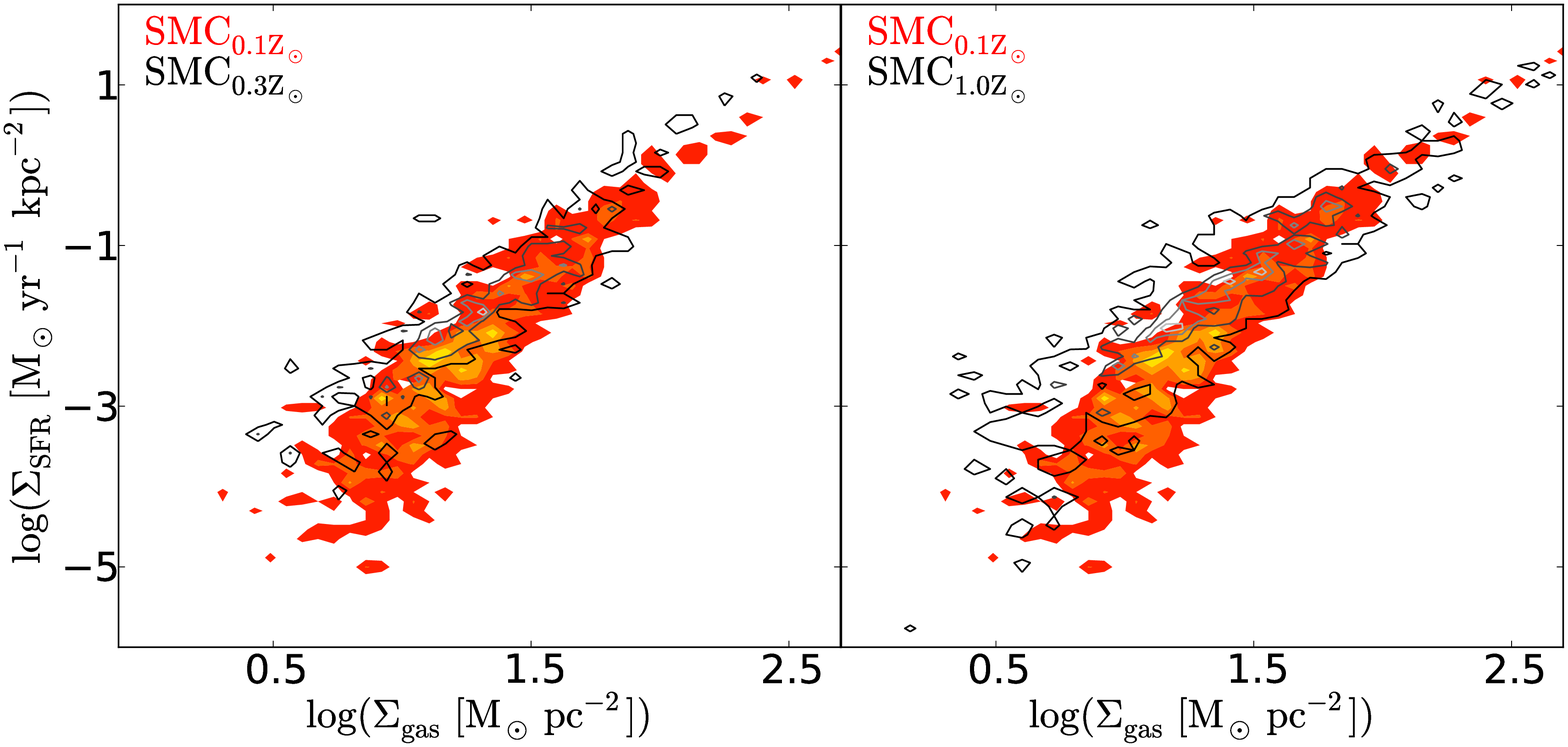}
\caption{\label{Fig:SMC_metallicity} The impact of gas metallicity on the \Sigmagas--\SigmaSFR{} relation in the model of \smc. The left panel shows two 
simulations of the \smc{} with comparable metallicities: 0.1 \Zsun (colored filled contours) and 0.3 \Zsun{} (black contours). The right panel compares the effect of gas
metallicity of 0.1 \Zsun{} with that of 1.0 \Zsun. Gas cooling rates increase with metallicity, which translates into increased \SigmaSFR{} for a fixed value 
of \Sigmagas{} in the region of the break. The color coding of contour levels is as in Figure \ref{Fig:MW_beam_compare_all}.}
\end{center}
\end{figure*}

Figure \ref{Fig:SMC_metallicity} shows the impact of metallicity on the \SigmaSFR. The left panel compares two systems with comparable
metallicities, 0.1 \Zsun{} and 0.3 \Zsun, while on the right panel, two more extreme metallicities are compared, 0.1 \Zsun{} and 1.0 \Zsun. In the region 
below the break, high metallicity systems tend to have higher \SigmaSFR{} for a fixed value of \Sigmagas{} than systems with lower metallicity.

\begin{figure*}[t!]
\begin{center}
\includegraphics[width=\textwidth]{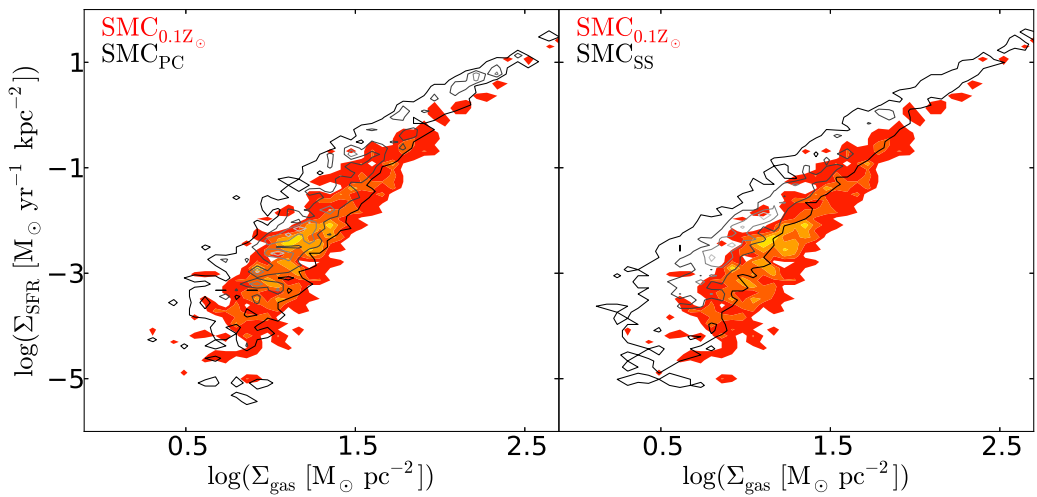}
\caption{\label{Fig:SMCs} Comparison of the \Sigmagas--\SigmaSFR{} relation in \smcZth{} (colored filled contours) to that of \smcpc{} (black contours) in 
                          the left panel and to that of \smcss{} (black contours) in the right panel). 
                          The color coding of contour levels is as in Figure \ref{Fig:MW_beam_compare_all}.}
\end{center}
\end{figure*}

The impact of the \eos{} on the SFR is presented in Figure \ref{Fig:SMCs}.
The \smcZth{} simulation is compared to that of \smcpc{}, using the \eos{} of pseudo-cooling and to that of \smcss{} with the \eos{} of 
self-shielding.   
The similarity of two contour plots on the left panel shows that the pseudo-cooling \eos{} is a good approximation to the actual heating and cooling processes
even for a slightly lower metallicity in this case (we remind that the pseudo-cooling \eos{} is derived using the metallicity of 
1/3 \Zsun{}; see Section \ref{Sec:EOS}). 
In the case of the self-shielding \eos, for a given value of \Sigmagas{}, \SigmaSFR{} tends to be higher compared to that of the simulation with metallicity 
of 0.1 \Zsun. 

We do not assume any metallicity gradient in the gas, nor chemical evolution. We use the model for self-shielding without an implicit metallicity dependence,
similarly to the work of \cite{Dobbs2008}. 
As shown in Figure \ref{Fig:SMCs}, the self-shielding EoS leads to higher \SigmaSFR{} for fixed \Sigmagas{} compared to the 
model of \smc{} with metallicity of 0.1 \Zsun.

The existence of the break in the Schmidt-Kennicutt relation in our models does not seem to depend on self-shielding effects. The exact position of this 
break is however sensitive to metallicity: the slope at low \Sigmagas, i.e. below the break, is higher 
in metal-poor galaxies as shown on Figure \ref{Fig:SMC_metallicity}.
Similar metallicity dependent position of the break is present in the theoretical model of \cite{Krumholz2009} including the effect of 
hydrogen self-shielding which in turn determines the amount of gas in molecular form. 
In addition, \cite{Dib2011} explored the metallicity-dependent feedback and found that it can lead to a modification of the position of the break for a 
given metallicity-dependent molecular gas fraction.
It is clear that self-shielding has an impact on the Schmidt-Kennicutt relation (see right panel of the Figure \ref{Fig:SMCs}), but it does not seem to be the 
only factor determining the presence of the break.

\subsection{Local parameters}
\label{Sec:Local}

\subsubsection{Mach number}
\label{Subsec:Mach_nmber}
\begin{figure}[t!]
\begin{center}
\includegraphics[width=\columnwidth]{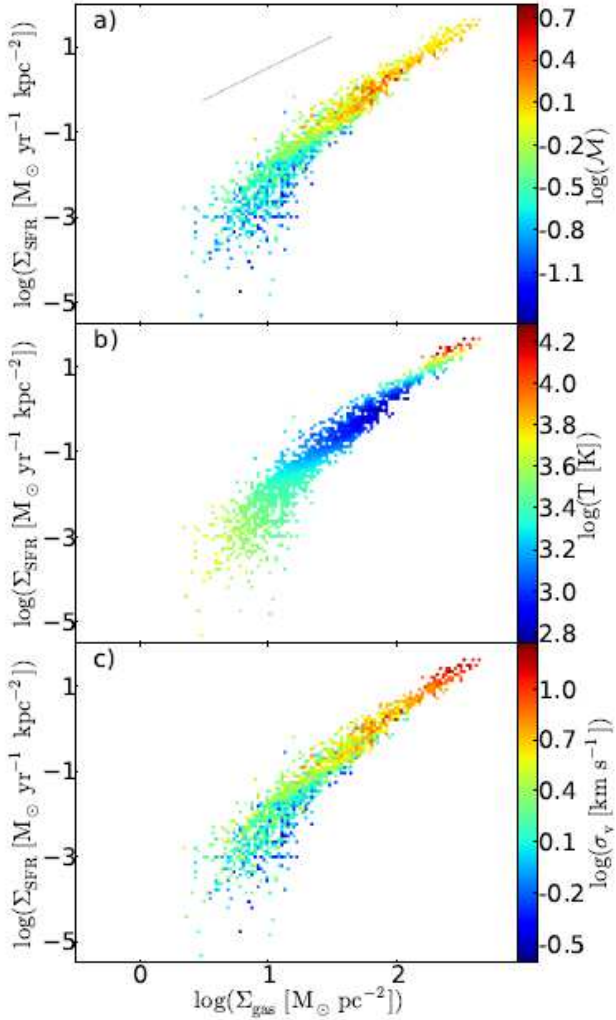}
\caption{\label{Fig:MW_Mach_sigma_T} 
The local surface density of the star formation rate as a function of surface density of gas for the \mwpc{} model. The color indicates the Mach number (panel a), 
temperature (panel b) and velocity dispersion (panel c) in each beam. The dotted line indicates a power-law of index $3/2$. Note that regions at high \Sigmagas{}
and high \SigmaSFR{} that have high temperatures represent unresolved dense gas situated on the Jeans polytrope (see Section \ref{Sec:EOS}).} 
\end{center}
\end{figure}

\begin{figure}[t!]
\begin{center}
\includegraphics[width=\columnwidth]{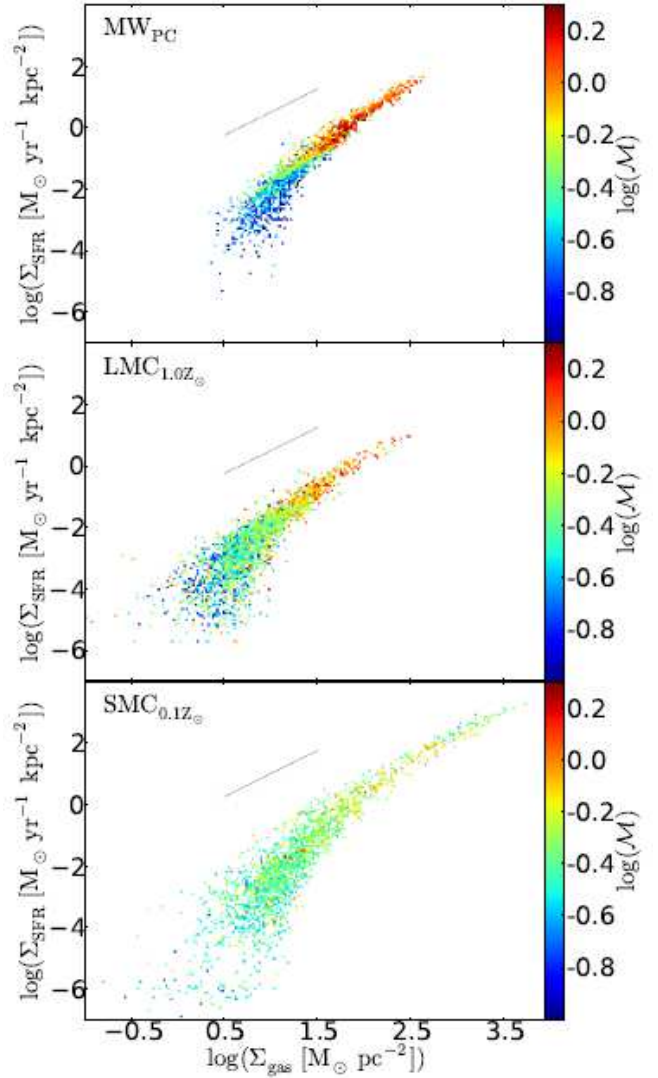}
\caption{\label{Fig:MW_SMC_LMC_Mach} 
The local surface density of the star formation rate as a function of surface density of gas for \mwpc{} (top panel), \lmcZsun{} (middle) and \smcZth{} (bottom). 
The color indicates the Mach number. The dotted line indicates a power-law of index $3/2$.
}
\end{center}
\end{figure}

In Figure \ref{Fig:MW_Mach_sigma_T} we show how the \Sigmagas{} -- \SigmaSFR{} relation depends on the Mach number, temperature and velocity dispersion calculated 
using the Equations \ref{Eq:Mach_number}, \ref{Eq:Temperature} and \ref{Eq:sigma_v}, respectively. We show the example of \mwpc, but we obtain 
qualitatively similar results for all other galaxies. The Mach number dependence for \mwpc, \smcZth{} and \lmcZsun{} is displayed in Figure \ref{Fig:MW_SMC_LMC_Mach}.

Two regimes in the star formation relation are identified. The points located in the region below the break have typically Mach numbers with values below unity. 
Furthermore, for a given \Sigmagas, \SigmaSFR{} increases with increasing Mach number.
At high surface densities of gas, \SigmaSFR{} and \Sigmagas{} are found to be correlated. The gas reservoirs that happen to be in this regime of efficient
star formation tend to have supersonic velocity dispersions.

Both the temperature and velocity dispersion contribute to the resulting Mach number dependence in the star formation relation. 
Despite the variation in temperature, the overall variation in Mach number relies on $\sigma_v$.
The velocity dispersion of the ISM can be increased by several processes. Among them \cite{Bournaud2010} found, in simulations similar 
to those analyzed here, self-gravity to play the dominant role, compared to stellar feedback.
Therefore, by increasing the velocity dispersion, self-gravity sets the level of turbulence, i.e. the compression of gas and thus the SF.
This suggests that the power-law part of the \Sigmagas--\SigmaSFR{} relation arises from self-gravity at high Mach number, while this 
connection is weaker in the break.

\subsubsection{Vertical scale of the gas}

\begin{figure}[h!]
\begin{center}
\includegraphics[width=\columnwidth]{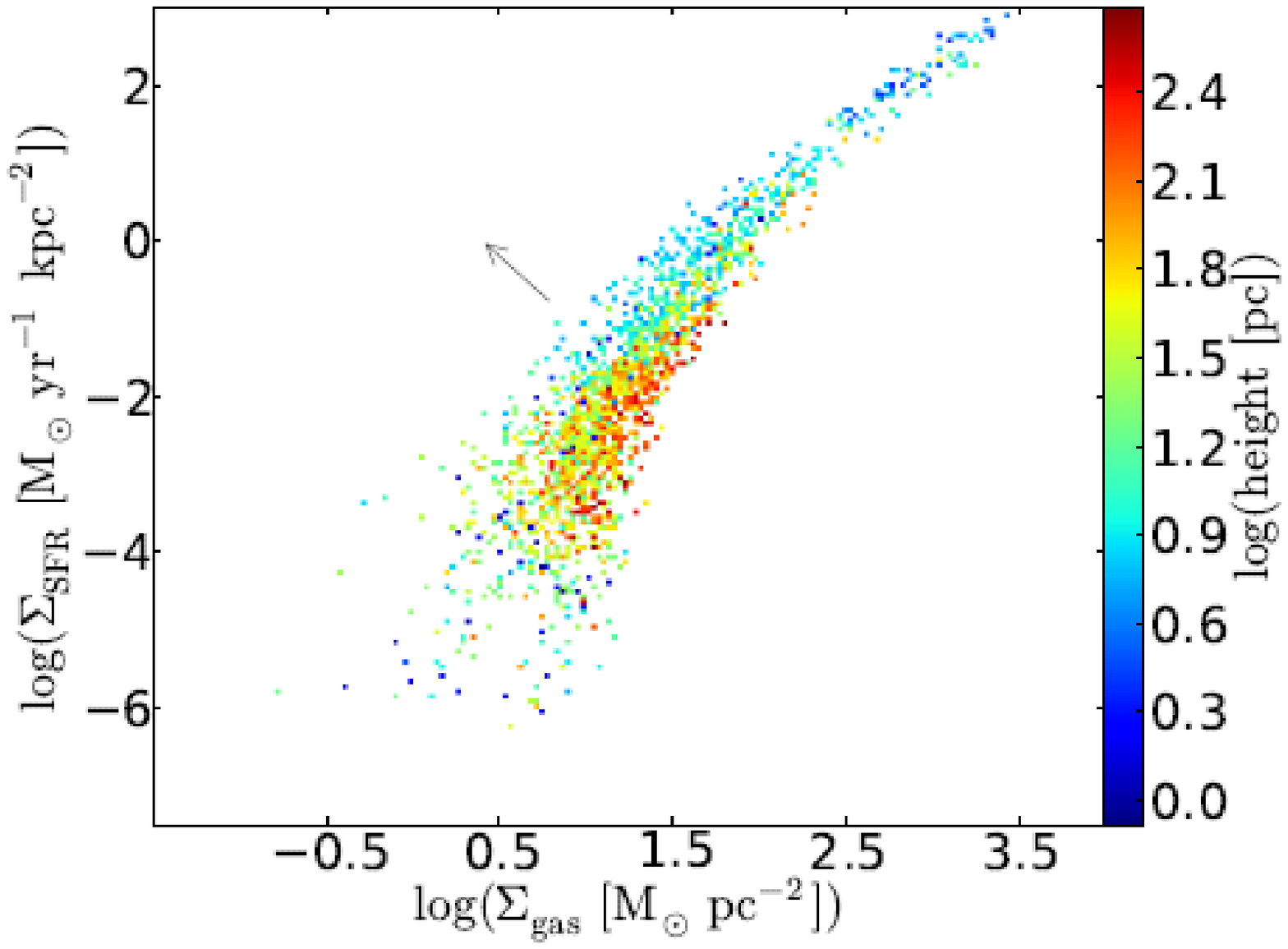}
\caption{\label{Fig:SMC_h} The local surface density of the star formation rate as a function of surface density of gas for the \smcZth{} model. 
The color represents the thickness of star-forming region. The arrow indicates the direction in the measured scale-height of the 
gas from higher to lower values.}
\end{center}
\end{figure}

Figure \ref{Fig:SMC_h} shows the variation of the \Sigmagas--\SigmaSFR{} relation with the thickness of the star-forming regions in \smcZth{}. 
For a given surface gas density, thicker regions tend to have lower surface star formation density. This relation between \SigmaSFR{} and 
the thickness parameter at fixed \Sigmagas{} results from the Equation \ref{Eq:SFR} relating the volume density of gas with that of star formation rate. 

\section{Discussion}
\label{Sec:Discussion}

\subsection{Comparison with observations}
\label{Sec:Comparison}

Most spatially resolved studies of spiral galaxies find the presence of a power-law \Sigmagas--\SigmaSFR{} relation with
a break at surface gas densities of the order of a few M$_{\odot} \, \mathrm{pc}^{-2}$ \citep[see][and references therein]{KennicuttEvans2012}. The slope of the 
power-law relation in the high surface-density regime is found to be in the range 1.2--1.6 when total (molecular plus atomic) gas surface density is considered.

Less agreement about the power-law slope in observations is reached when molecular-gas surface density is considered solely. 
Some recent studies \citep[e.g.][]{Eales2010,Rahman2011,Leroy2013} have reported an
approximately linear relation between the surface density of star formation rate and the surface density of molecular gas. Other 
studies \citep[e.g.][]{Kennicutt2007,Verley2010,Liu2011} have found a much steeper relation, with a slope in the range 1.2--1.7, similar to that of integrated 
measurements \citep{Kennicutt1998}. This discrepancy between different results in observations is still debated. A possible interpretation 
of the sublinear relation was recently proposed by \cite{Shetty2013}. They suggest that the CO emission used in the estimation of \Sigmagas{} is not all
necessarily associated with SF. Not subtracting off such a diffuse component could lead to a slope close to unity.

The distribution of data points from the observations of the \smc{} \citep{Bolatto2011} in the \Sigmagas--\SigmaSFR{} plane has a similar shape than
that of spiral galaxies, but is noticeably shifted toward higher total \Sigmagas.

\begin{figure}[h!]
\begin{center}
\includegraphics[width=\columnwidth]{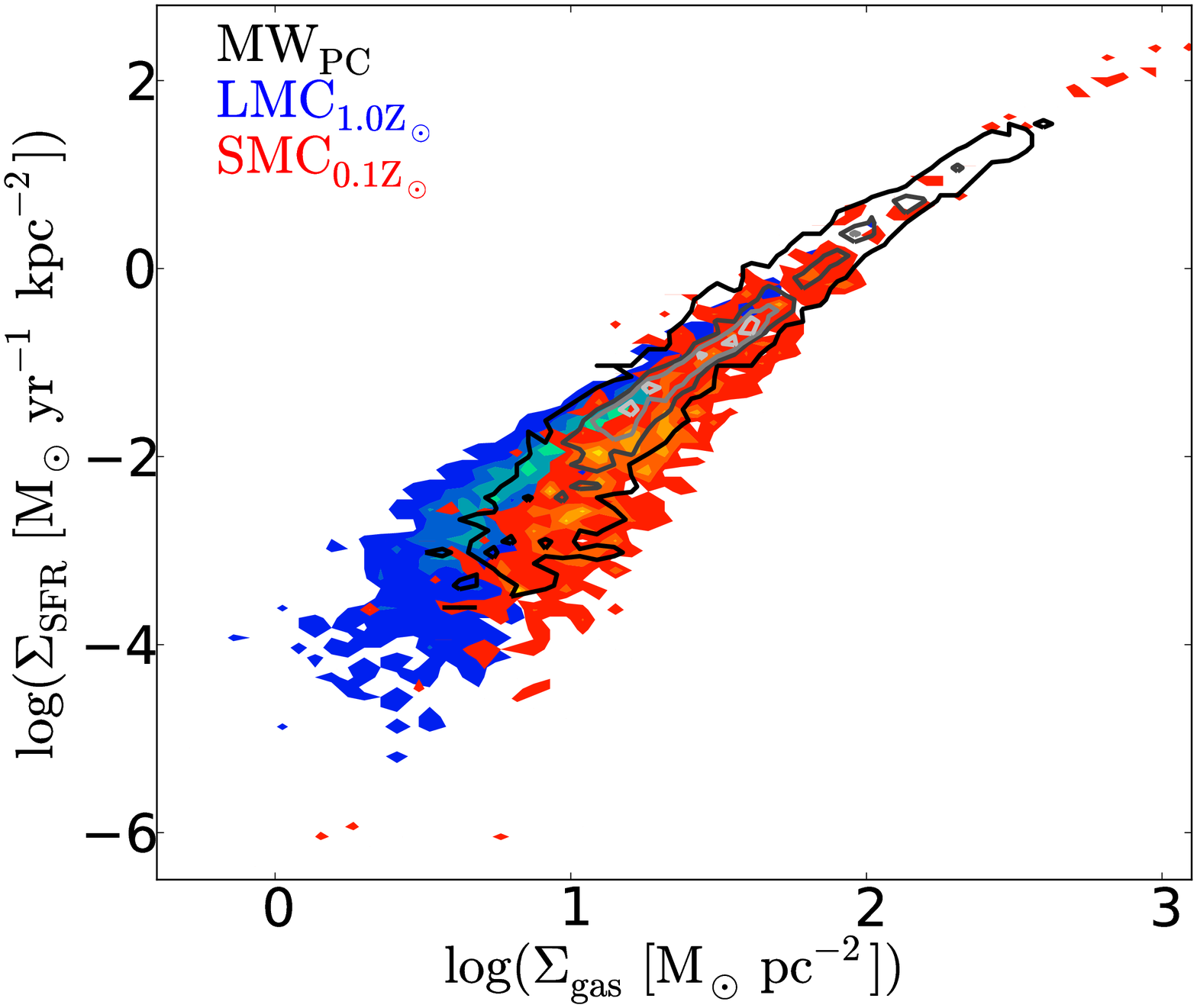}
\caption{\label{Fig:contour_all} Comparison of \mwpc, \lmcZsun{} and \smcZth. The color coding of contour levels is as in Figure \ref{Fig:MW_beam_compare_all}.}
\end{center}
\end{figure}

In Figure \ref{Fig:contour_all}, we show three of our models: \mwpc, \smcZth{} and \lmcZsun{} in the \Sigmagas--\SigmaSFR{} plane. 
The \mwpc{} and the \lmcZsun{} models lie in the loci of observed spiral galaxies \citep[e.g.][]{Kennicutt1998,Kennicutt2007,Bigiel2008}.
Our \smcZth{} model has a lower \SigmaSFR{} for a given \Sigmagas{} when compared to both the \mwpc{} and the \lmcZsun{} models. 
The region below the break of our \smcZth{} model is located at slightly lower \Sigmagas{} than the real Small Magellanic Cloud, 
but its displacement with respect to spiral galaxies (\mw{} and \lmc) is well reproduced (Figure \ref{Fig:contour_all}). 
In our simulations, Equation \ref{Eq:SFR} sets the slope of power-law relation with the value of 1.5.
A shallower relation, closer to the observed values, might be reached by accounting for a stronger regulation of star formation 
(e.g. pre-SN stellar feedback, see \citealt{Renaud2012}), but this slope change has not been demonstrated by simulations yet.

\subsection{Interpretation of threshold}
The \textit{existence} of the break in the \Sigmagas--\SigmaSFR{} relation is, in our models, equivalent of having a non-zero value of the volume density 
threshold in the local, three dimensional star formation relation. Setting no threshold leads to a power-law relation without a break. 

\begin{figure}[t!]
\begin{center}
\includegraphics[width=\columnwidth]{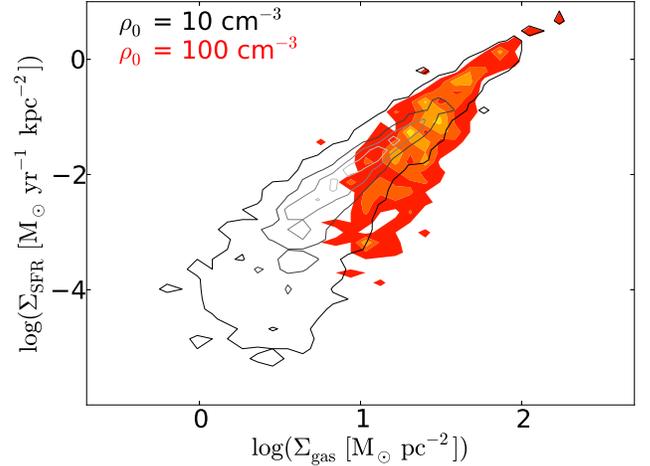}
\caption{\label{Fig:th_LMC} Comparison of the \Sigmagas--\SigmaSFR{} relation in the \lmcZsun{} simulation with the star formation volume density threshold
$\rho_0$=10 \Hcc{} (black contours) to that with $\rho_0$=100 \Hcc{} (colored filled contours). The color coding of contour levels is as in 
Figure \ref{Fig:MW_beam_compare_all}.}
\end{center}
\end{figure}

Figure \ref{Fig:th_LMC} shows that the value of the density threshold $\rho_0$ that we have used in our analysis has an impact on the \Sigmagas--\SigmaSFR{} 
relation. Changing the value of $\rho_0$ changes the slope at low \Sigmagas{} in the \Sigmagas--\SigmaSFR{} relation (Figure \ref{Fig:th_LMC}).
This could suggest that the transition from the inefficient to the power-law regime could be due to the density threshold $\rho_0$ we imposed by hand in the 
star formation law (see Equation \ref{Eq:SFR}). 
However, we have checked that the deviation from the power-law regime occurs at $\mathcal{M} \approx$ 1, independently of $\rho_0$.
In addition, in Figure \ref{Fig:MW_SMC_LMC_Mach}, we have shown that beams located in the break tend to have $\mathcal{M}$ below unity, while regions at high 
\Sigmagas{} are mostly supersonic. 

\begin{figure}[h!]
\begin{center}
\includegraphics[width=\columnwidth]{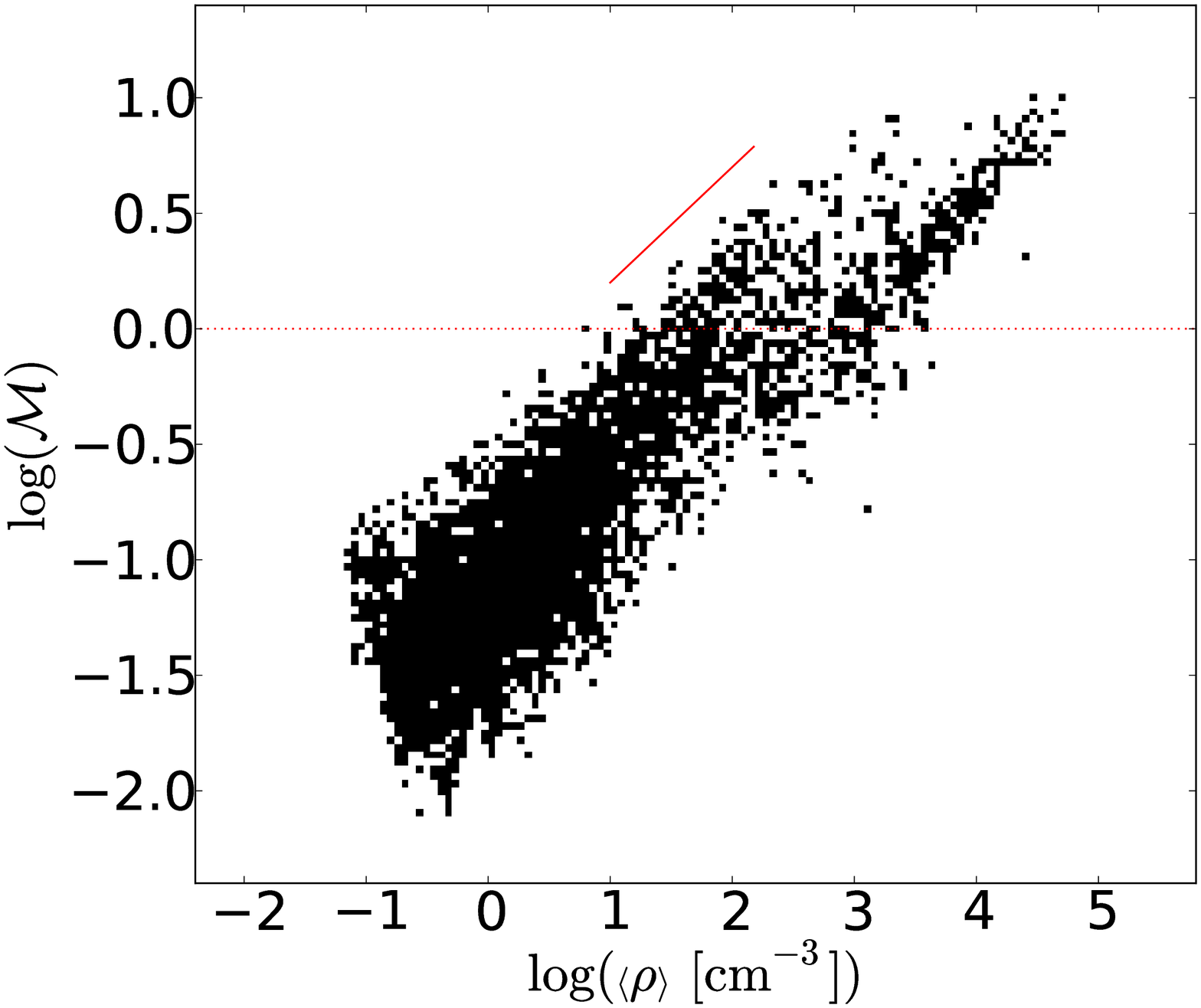}
\caption{\label{Fig:Mach_rho_T} The Mach number as a function of the average volume density of gas computed in beams of 100$\times$100$\times$100 pc$^3$, for \mwpc. 
The solid line corresponds to $\mathcal{M} \propto \left<\rho\right>^{0.5}$ \cite[][Figure 9]{AuditHenne2010}.}
\end{center}
\end{figure}

To better understand the behavior of the ISM in our simulations, we show in Figure \ref{Fig:Mach_rho_T} the Mach number as a function of average 
volume density of gas\footnote{computed as the mass-weighted average density of the gas in each beam} $\left<\rho\right>$ in the beam for \mwpc. The Mach number varies with the 
average density as $\mathcal{M} \propto \left<\rho\right>^{0.5}$, similarly to the two-phase turbulent flow studied by 
\cite{AuditHenne2010}. Although caution should be used when doing such comparison (temperature and velocity dispersion vary with density differently in both 
models), the onset of the supersonic regime, i.e. the transition from an inefficient regime to a power-law, happens at densities of $\approx 10$ \Hcc
\cite[see also][their Figures 4 and 9]{AuditHenne2010}. 

\begin{figure}[h!]
\begin{center}
\includegraphics[width=\columnwidth]{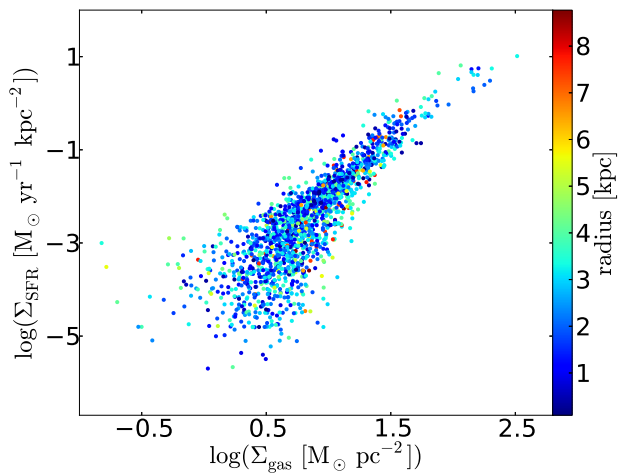}
\caption{\label{Fig:LMC_KS_radius2} The local surface density of the star formation rate as a function of surface density of gas for \lmcZsun. 
Color indicates the radial distance of the beam in kpc.}
\end{center}
\end{figure}

Other interpretations of the observed break are possible. The \Sigmagas--\SigmaSFR{} relation could be an effect of the galactic radial distance with low 
\Sigmagas{} at large radius and
high \Sigmagas{} at small radius, as found by \cite{Kennicutt2007} and \cite{Bigiel2008}. The break could then be explained as a consequence of the drop in 
the average volume density in the outer regions of galaxies as proposed by \cite{Barnes2012}. However, Figure \ref{Fig:LMC_KS_radius2} shows no such radial 
dependence for \Sigmagas{} nor \SigmaSFR. Star-forming regions in outer parts of a galaxy can exhibit both star formation regimes. A possible explanation why we 
do not see any radial dependence in our simulations may be a missing metallicity gradient. The outer regions of our simulated galaxies have the same metallicity 
than the innermost regions, thus the metallicity is probably too high at the edge of disk and allows for an efficient cooling and consequently an efficient 
star formation while it may lie in the break regime otherwise.

\begin{figure}[h!]
\begin{center}
\includegraphics[width=\columnwidth]{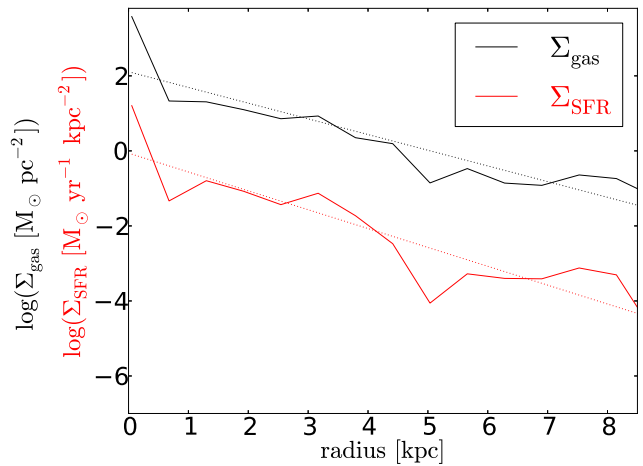}
\caption{\label{Fig:LMC_profil} Radial profiles of azimuthally averaged \Sigmagas{} (black) and \SigmaSFR{} (red) for \lmcZsun{}. Dotted lines correspond 
to exponential fits.}
\end{center}
\end{figure}

When azimuthally averaged, \Sigmagas{} and \SigmaSFR{} both decline steadily as a function of radius in many galaxies despite different morphologies 
(see \citealt{Bigiel2008} for a sample of nearby spiral galaxies and  \citealt{Leroy2009} for CO intensity radial profiles for the same sample). 
Figure \ref{Fig:LMC_profil} shows \Sigmagas{} and \SigmaSFR{} as functions of radius for \lmcZsun{}. 
Both radial profiles decline with galactic radius as in observed spiral galaxies.

Another alternative explanation for the existence of the break is that it corresponds to the transition from atomic to molecular hydrogen \citep{Krumholz2009}. 
According to this scenario, the transition from atomic to molecular hydrogen and the subsequent star formation depend on local conditions that vary with 
galactic radius, e.g. metallicity, gas pressure and shear\footnote{Shear dictates whether molecular clouds can form \citep[e.g.][]{Leroy2008,Elson2012}, 
but if they do form, shear does not seem to influence the efficiency at which they convert their gas into stars \citep{Dib2012}.}. \cite{Bigiel2008} found such radial dependence in the sample of
nearby galaxies in agreement with the findings of \citet{WongBlitz2002} and the threshold interpretations of e.g. \citet{Kennicutt1989}, \citet{MartinKennicutt2001} and \citet{Leroy2008}. 
Similar results are reproduced in some simulations, e.g. \cite{Halle2013}, who find that
molecular gas is a better tracer of star formation than atomic gas and plays an important role in the low surface density regions of galaxies by allowing for
more efficient star formation.
However, our models that do not include chemodynamics, are able to reproduce the observed break at low $\Sigmagas$. Therefore, this seems to indicate that 
the presence of molecules 
is not a necessary condition to \textit{trigger} the process of star formation. However, we acknowledge numerous observational evidences showing that
molecules are involved at a later stage of the SF process.

\begin{figure*}[t!]
\begin{center}
\includegraphics[width=\textwidth]{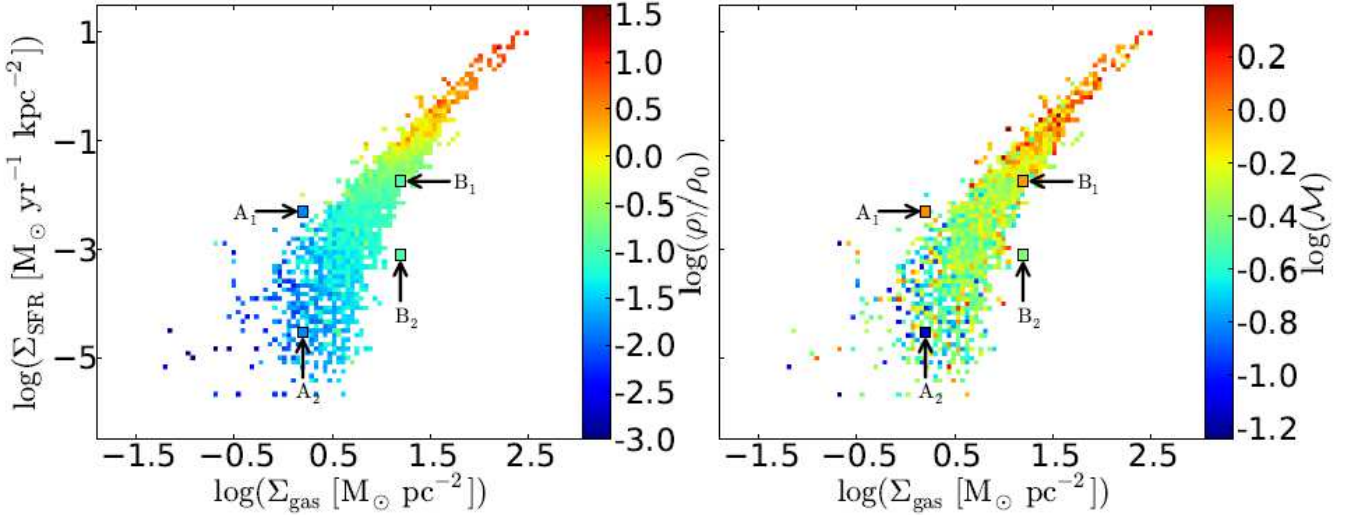}
\caption{\label{Fig:LMC_rho_rho0_Mach} The volume density normalized to the threshold density $\rho_0$=10 \Hcc{} (left panel) and the Mach number
(right panel) for the \lmcZsun{} simulation, as in Figure \ref{Fig:MW_SMC_LMC_Mach}. 
Two pairs of beams (A$_1$--A$_2$ and B$_1$--B$_2$) are highlighted in each panel. Within each pair, the two beams are chosen to have the same value of 
\Sigmagas{} and a similar value of $\left<\rho \right>/ \rho_0$. Beams with higher \SigmaSFR{} at fixed \Sigmagas{} have a higher Mach number. The two pairs of
beams have larger size for clarity reasons.}
\end{center}
\end{figure*}

To summarize, we consider two representative beams having the same \Sigmagas, but different \SigmaSFR{} 
(Figure \ref{Fig:LMC_rho_rho0_Mach}). 
These beams have similar average volume densities $\left<\rho\right>$ which can be several orders 
of magnitude smaller than $\rho_0$.
However, the beam that happens to have the highest \SigmaSFR{} has always the highest Mach number, as previously suggested by Figure \ref{Fig:MW_SMC_LMC_Mach}.
We have argued above that the density threshold $\rho_0$, the thickness of the star-forming regions and the molecules do not 
have impact on the transition from the regime of inefficient star formation to the efficient power-law regime.
The role of the artificial threshold $\rho_0$ imposed in the
simulations is to set a frontier between the diffuse non-star-forming gas and the star-forming component, but not to tune the efficiency of star formation per se.
Therefore at a given \Sigmagas, this efficiency depends mostly on the level of turbulence ($\mathcal{M}$), i.e. the compression of the ISM.

\begin{figure}[h!]
\begin{center}
\includegraphics[width=\columnwidth]{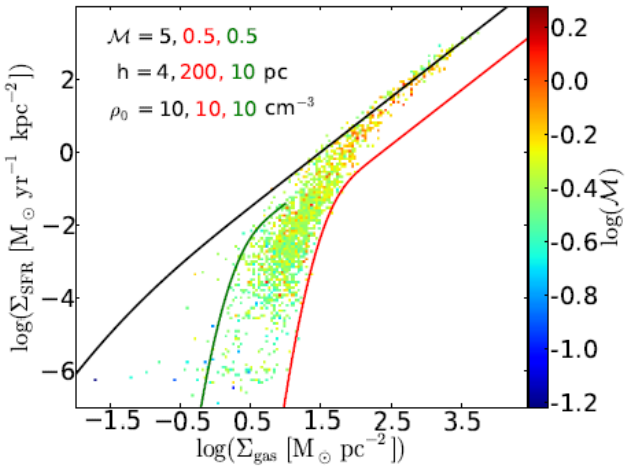}
\caption{\label{Fig:SMC_model} \smcZth{}: comparison with the \cite{Renaud2012} model with three sets of parameters (indicated in the legend). The black curve 
matches the supersonic regime of efficient star formation, while the green and the red curves represent upper and lower limits for the regime of the break, 
where the star formation is inefficient. The $h$ parameter is in agreement with values measured in the simulation (see Figure \ref{Fig:SMC_h}).}
\end{center}
\end{figure}

\begin{figure}[h!]
\begin{center}
\includegraphics[width=\columnwidth]{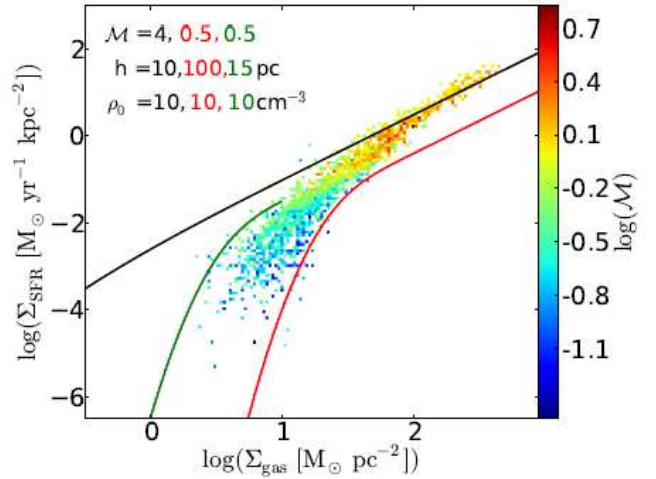}
\caption{\label{Fig:MW_model} \mwpc{}: comparison with the \cite{Renaud2012} model. As in Figure \ref{Fig:SMC_model}, the supersonic regime is compared to the model
prediction and similarly, the subsonic regime at low \Sigmagas{} is situated between the curves characterized by the Mach number lower than unity for the 
measured thickness.}
\end{center}
\end{figure}

\cite{Renaud2012} proposed that the break indeed corresponds to the onset of supersonic turbulence which, by generating shocks, triggers gravitational instabilities 
leading to star formation. 
In Figures \ref{Fig:SMC_model} and \ref{Fig:MW_model}, we compare simulations of \mwpc{} and \smcZth{} with the analytical model of \cite{Renaud2012}. In this 
model, the relation between \Sigmagas{} and \SigmaSFR{} depends on three parameters: the Mach number $\mathcal{M}$, the scale-height $h$ and the star formation 
volume density threshold $\rho_0$. 
We do not compare each star-forming region in the simulation with the model, but we are rather interested in what 
values these parameters should take to obtain upper and lower limits for simulated data. 
The break is in the subsonic regime (measured values of $\mathcal{M}$ are below unity) which corresponds to the regime where the analytical model deviates from 
its asymptotic behavior (at high \Sigmagas). In this regime the scale-heights of the beams set the efficiency of star formation spanning the range given by 
the model and quantitatively in accordance with the values measured in the simulations (Figure \ref{Fig:SMC_h}).
In the analytical model, the power-law regime can be reached even with the Mach number below unity (red curve). However, our simulations do not probe this area of the 
\Sigmagas--\SigmaSFR{} plane: the data points in the power-law regime are exclusively supersonic and can only be described by a model with the Mach number
above unity (black curve).

\subsection{Metallicity}

\begin{figure*}[t!]
\begin{center}
\includegraphics[width=\textwidth]{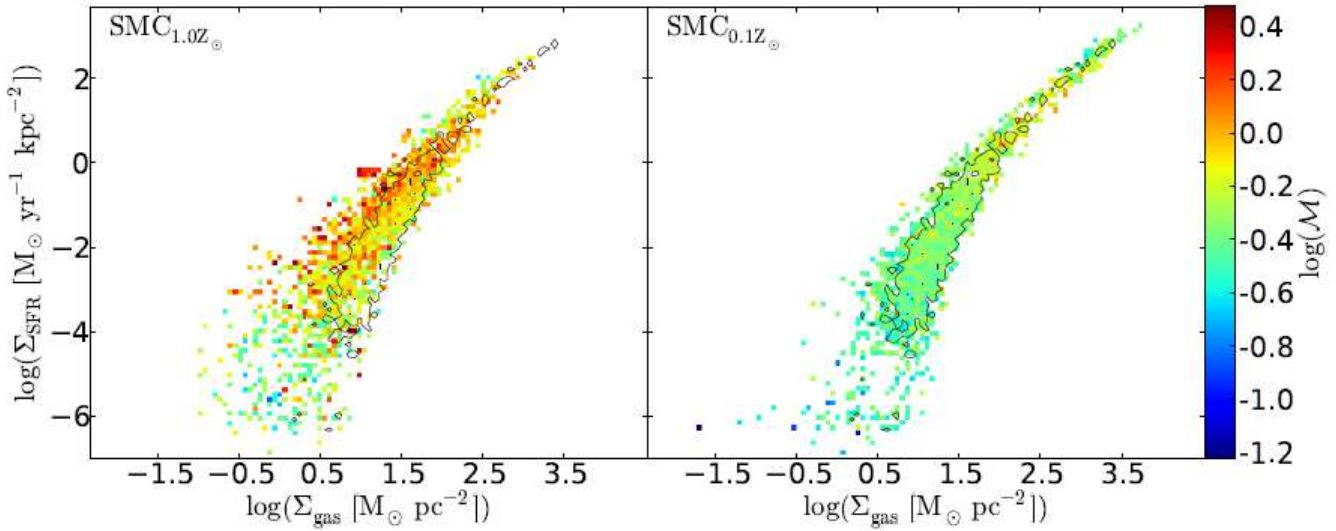}
\caption{\label{Fig:SMC_metallicity2} As in Figure \ref{Fig:SMC_metallicity}, the effect of gas metallicity on the \Sigmagas--\SigmaSFR{} relation in the model 
of \smc{} is represented. Colors indicate the Mach number in the simulation of the \smc{} with metallicity of 1.0 \Zsun{} on the left panel and  
the simulation of the \smc{} with metallicity of 0.1 \Zsun{} on the right panel. In both panels the black contours are those of the simulation of the 
\smcZth, shown for reference. }
\end{center}
\end{figure*}

In Figure \ref{Fig:SMC_metallicity}, we have shown that the exact position of the break in the \Sigmagas--\SigmaSFR{} plane depends on metallicity. 
A comparison of different metallicities in otherwise identical systems shows that the slope at low \Sigmagas{} has a greater value in metal-poor galaxies. 
Figure \ref{Fig:3pdf} suggests that metallicity is not the only factor determining the gas density distribution in our simulations. Similar 
lack of direct dependence of the fraction of dense gas on metallicity is found when simulations of the \smc{} with different metallicities are
compared (not shown here). Thus the slope at low \Sigmagas{}
cannot be explained by the presence of a higher fraction of dense gas in systems with higher metallicity compared to systems with lower 
metallicity. However, metallicity has an impact on star formation, even though indirect. Metallicity directly influences the temperature
of the gas: higher the metallicity, more efficient the cooling therefore the temperature and, in turn, impacts the 
Mach number. In Figure \ref{Fig:SMC_metallicity2}, we show Mach numbers for \smcZsun{} and \smcZth{}.
Higher values of Mach number are reached in galaxy with higher metallicity.

This work does not include metallicity-dependent self-shielding and feedback. Accounting for them, 
\cite{Dib2011} showed that both the fraction of gas in molecular form and the efficiency of star formation per unit time depend on metallicity. This leads to 
the metallicity dependent \Sigmagas--\SigmaSFR{} relation at any \Sigmagas. 

\section{Summary}

\label{Sec:Summary}
In this paper, we study the star formation relations and thresholds at 100 pc scale in a sample of low-redshift simulated galaxies. 
These include simulations representative of Milky Way-like spiral galaxy, the Large and the Small Magellanic Clouds.
We analyze the role of interstellar turbulence, gas cooling, and geometry in drawing these relations, by investigating the dependence of the star formation  
on three parameters: the Mach number, the thickness of the star-forming region and the star formation volume density threshold.
We compare the simulated data with the idealized model for star formation of \cite{Renaud2012}.

Our main findings are as follows:
\begin{enumerate}
 \item Our simulations support an interpretation of the surface density threshold for efficient star formation as the typical density for the onset of 
       supersonic turbulence in dense gas, 
       as proposed theoretically by \cite{Renaud2012}. For all analyzed systems, we obtain qualitatively the same result: regions located below the break are 
       dominated by subsonic turbulence, while turbulence tends to be supersonic in those located in the power-law regime, . 
 \item The distribution of the ISM of a galaxy in the \Sigmagas--\SigmaSFR{} plane (mainly the position of the break) is sensitive to metallicity, but always 
       correlated with the Mach number as detailed above. When different metallicities are considered for otherwise identical systems, 
       \SigmaSFR{} increases with the metallicity. When different systems with same metallicities are considered (compare Figure \ref{Fig:contour_all}
       for \lmcZsun{} and Figure \ref{Fig:SMC_metallicity} for \smcZsun), roughly the same position in the \Sigmagas--\SigmaSFR{} plot is 
       obtained. This can explain observations of low-efficiency star formation in relatively dense gas in SMC-like dwarf galaxies. 
       The driving physical parameter is still the onset of supersonic turbulence, but this onset is harder to reach at moderate gas densities in 
       lower-metallicity systems that can preserve warmer gas.
 \item The vertical spread in the \Sigmagas--\SigmaSFR{} plot is given by the interplay between different parameters of star-forming regions. 
       Figures \ref{Fig:SMC_model} and \ref{Fig:MW_model} show a reasonable agreement between simulations and the analytic model of \cite{Renaud2012}, 
       confirming that this idealized model provides a viable description of star formation in a turbulent ISM compared to more realistic simulations of 
       self-gravitating systems with star formation and feedback.
       The values of the model parameters (Mach number, thickness and density threshold) characterizing the points in \Sigmagas--\SigmaSFR{} plane
       are close to the values measured in simulations.
\end{enumerate}

Several other models \citep[e.g.][]{Krumholz2009} have proposed that self-shielding alone is efficient at producing giant molecular clouds and 
triggering SF. 
Indeed, this effect cools the gas down at high density, thus enhancing the fragmentation of the ISM,
but also lowering the sound speed, i.e. increasing the level of turbulence. Both the compression of the ISM by supersonic turbulence and the 
fragmenting effect from self-shielding increase with metallicity. Having neglected the dependence of self-shielding on metallicity,
our results emphasize only the role of supersonic turbulence in our most metal rich examples. Combining the two effects would lead to a higher 
efficiency of star formation than either effect alone.

At the scale of clouds, the gravitational collapse is known to trigger SF. However, at larger scales, in
galactic structures like spiral arms, we found that the injection of turbulence by self-gravity (and possibly by other processes like shear 
and feedback) can drive the compression of the gas, leading to SF. 
In this view, an external trigger like supersonic turbulence could be a sufficient condition to from stars, without necessarily invoking the
collapse of large galactic regions ($\sim$ 100 pc) prior to turbulent compression -- only compressed regions need to eventually collapse into stars.


\acknowledgments

We thank the anonymous referee for suggestions which improved the paper.
The simulations presented in this work have been performed at the TGCC (France) under GENCI (04-2192) and PRACE allocations.
FR and FB acknowledge support from the EC through grant ERC-StG-257720.

\end{document}